\documentclass[journal,12pt,onecolumn,draftclsnofoot]{IEEEtran}
\usepackage{epsfig}
\usepackage{amsmath}
\usepackage{amsmath}
\usepackage{mathrsfs}
\usepackage{epsfig}
\usepackage{amsfonts}
\usepackage{amssymb}
\usepackage{cite}
\usepackage{tabularx}
\usepackage{mathrsfs}
\usepackage{euscript}
\usepackage{graphicx}
\usepackage{multirow}
\usepackage{stfloats}
\usepackage{array}
\usepackage{amsmath}

\usepackage{cite}
\usepackage{color}
\usepackage{bm}
\usepackage{subfigure}
\usepackage[]{algorithm2e}
\usepackage{amsthm}
\newtheorem{prop}{Proposition}
\begin{document}
\makeatletter
\renewcommand{\@algocf@capt@plain}{above}
\makeatother
\title{Dynamic Non-Orthogonal Multiple Access (NOMA) and Orthogonal Multiple Access (OMA) in 5G Wireless Networks}
\author{
\authorblockN{Mina Baghani$^1$, Abbas Mohammadi$^1$, Mahdi Majidi$^1$, Mikko Valkama$^2$\\}
\thanks{$^{(1)}$ Microwave and Wireless Communication Research Lab., Electrical Engineering Department, Amirkabir University of Technology, Tehran, Iran\\$^{(2)}$
Department of Electronics and Communications Engineering, Tampere University of Technology, Tampere, Finland }
}
\author{
\small
    \IEEEauthorblockN{ Mina Baghani\IEEEauthorrefmark{1}, Saeedeh Parsaeefard\IEEEauthorrefmark{1}, Mahsa Derakhshani\IEEEauthorrefmark{2}, and Walid Saad \IEEEauthorrefmark{3}}\\
    \IEEEauthorblockA{\IEEEauthorrefmark{1} Communication Technologies \& Department, ITRC, Tehran, Iran} \\
    \IEEEauthorblockA{\IEEEauthorrefmark{2}Wolfson School of Mechanical, Electrical \& Manufacturing Engineering,  Loughborough University, UK}\\
    \IEEEauthorblockA{\IEEEauthorrefmark{3}Wireless@VT, Electrical and Computer Engineering Department, Virginia Tech, VA, USA}
    \vspace{-12mm}
    }




\maketitle

\begin{abstract}

In this paper, facilitated via the flexible software defined structure of the radio access units in 5G, we propose a novel dynamic multiple access technology selection among orthogonal multiple access (OMA) and non-orthogonal multiple access (NOMA) techniques for each subcarrier. For this setup, we formulate a joint resource allocation problem where a new set of access technology selection parameters along with power and subcarrier are allocated for each user based on each user's channel state information. Here, we define a novel utility function taking into account the rate and costs of access technologies. This cost reflects both the complexity of performing successive interference cancellation and the complexity incurred to guarantee a desired bit error rate. This utility function can inherently demonstrate the trade-off between OMA and NOMA. Due to non-convexity of our proposed resource allocation problem, we resort to successive convex approximation where a two-step iterative algorithm is applied in which a problem of the first step, called access technology selection, is transformed into a linear integer programming problem, and the nonconvex problem of the second step, referred to power allocation problem, is solved via the difference-of-convex-functions (DC) programming. Moreover, the closed-form solution for power allocation in the second step is derived. For diverse network performance criteria such as rate, simulation results show that the proposed new dynamic access technology selection outperforms single-technology OMA or NOMA multiple access solutions.

\end{abstract}


\begin{IEEEkeywords}
Orthogonal multiple access (OMA), non-orthogonal multiple access (NOMA), technology selection.
\end{IEEEkeywords}

\IEEEpeerreviewmaketitle
\section{Introduction}

The anticipated exponential growth in the demand for wireless access is expected to strain the capacity and coverage of existing wireless cellular networks \cite{exponen1,exponen2,sad1}. In particular, the fixed multiple access techniques of yesteryears, such as time division multiple access (TDMA), code division multiple access (CDMA), and frequency division multiple access (FDMA), which guarantee the orthogonality in time, code and frequency, respectively, will no longer be able to sustain this growing demand for wireless access. In order to address this challenge in the fifth generation (5G) of cellular systems, several new techniques for multiple access have recently emerged based on the concept of non-orthogonal multiple access (NOMA) as discussed in~\cite{nomafirst,sadA,sadB,scma,pdma,secnoma}.
In power-domain NOMA, multiple users can share each subcarrier and the diversity on that subcarrier is obtained by allocating different power levels to the users.
The basic principle of NOMA is to exploit the difference in channel gains among users in order to offer multiplexing gains. For example, in a two-user NOMA case, the lower power level is allocated to the user with higher channel gain (the first user) compared to the user with lower channel gain (the second user). Then, the information of different users is superimposed and transmitted.


Despite the proven benefits of NOMA \cite{advannoma2,advannoma3,advannoma}, several practical challenges must be addressed before NOMA can be effectively deployed. One such challenge is to analyze the sensitivity of NOMA to the accuracy of channel state information (CSI) \cite{csinoma}. Another major challenge is the complexity of transceivers.
Indeed, a typical NOMA transceiver requires the use of superposition coding and successive interference cancellation (SIC). Moreover, the performance of NOMA can be substantially limited when the difference in the channel gains of the involved users is not sufficiently significant. Clearly, these practical issues make it challenging to solely rely on NOMA, particularly, when the wireless users experience somewhat similar channel gains.

In particular, the performance of NOMA degrades when the difference in channel gains among the wireless users is small. Therefore, the need for a more complex receiver coupled with the higher probability of error imposed by SIC might limit the practicality of using NOMA in 5G under all network conditions.
One promising approach to overcome this issue is to leverage the software-defined nature of 5G systems \cite{sdr}, in order to implement a dynamic approach for multiple access selection depending on the network state, e.g. CSI.
This motivates the development of new multiple access solutions that can dynamically select between NOMA and OMA such as orthogonal frequency division multiple access (OFDMA) \cite{lte}.



The main contribution of this paper is a new framework for multiple access technology selection that can enable a 5G system to flexibly decide on whether to use NOMA or OMA depending on the state of its users. This programmable structure can be implemented in practical systems by using the inherent software defined structure of 5G and beyond \cite{sdr}.
The problem is formulated as an optimization problem whose objective
captures the tradeoff between the achievable rate and a processing cost of using each access technology.
Considering the processing cost of NOMA, the access technology selection between OMA and NOMA for each subcarrier is then used as an example of a dynamic, selective access scheme. As a result, a new utility function is proposed and defined as the total rate of network minus the cost of performing NOMA for the allocated subcarriers. Furthermore, to enhance infrastructure utilization, we consider a virtualized network in which each service provider (SP) has its own quality-of-service (QoS) requirements, which should be guaranteed via effective resource allocation \cite{virtu}.

The formulated problem is then shown to be nonconvex and complex to solve. To address this challenge, we propose a two-step iterative algorithm. To this end, the variables of the resource allocation problem are divided into two groups and are optimized iteratively in two steps to alleviate the problem complexity \cite{update}. The problem of allocating the first group of variables is transformed into a linear integer programming problem. To solve the power allocation in the second problem, DC programming is applied \cite{dc}. For the power allocation strategy, the closed-form solution is also derived. The obtained expression sheds light on the effects of the NOMA processing costs on the power allocation strategy.
To study the performance of our proposed resource allocation strategy, three criteria are investigated in simulation section rate, utility and outage probability. The outage probability is defined as the probability of not meeting constraints of optimization problem simultaneously.
Simulation results show that the proposed dynamic multiple access selection approach yields significant performance gains in terms of achievable rate, utility value, and outage probability compared to single-technology OMA or NOMA multiple access techniques.
In particular, when the maximum power limitation or minimum required rate of SPs is a dominant constraint of feasibility, our proposed scheme achieves 20\% higher utility performance compared to the cases in which pure OMA and NOMA are adopted.

\subsection{Related Works}


There exists a large body of works that addressed the resource allocation problems for OFDMA \cite{omasur,omasur2,omasur3,omasur4} and NOMA \cite{nphard,ratenoma,eenoma,varobj,lowcom} techniques in 5G. In OFDMA setup, the proposed resource allocation problems are consisting of subcarrier and power allocation. Meanwhile, in NOMA, one needs to optimize user pairing along with power and sub-carrier allocation \cite{ratenoma}. For both NOMA and OFDMA, network optimization is often posed using non-deterministic polynomial-time (NP)-hard problems \cite{nphard,npofdm}.


The authors in \cite{varobj} studied the problem of analytically characterizing the optimal power allocation for NOMA, considering various objective functions and constraints. To reduce the computational complexity, a new user pairing and power allocation scheme is proposed in \cite{lowcom}.
Meanwhile, the works in \cite{duplexnoma,mimonoma,het} have studied advanced approaches that combine NOMA with other emerging transmission techniques, such as full-duplex communications and multiple-input, multiple-output (MIMO) systems and heterogeneous systems.
	
This idea of utilizing a hybrid of OMA and NOMA in 5G has been recently studied in \cite{new} and \cite{new2}. In \cite{new}, the access technology is predetermined per each region consisting of four users. Despite being interesting, this work does not optimize the access technology selection while taking into account the instantaneous CSI.
In \cite{new2}, a heterogeneous network in which OMA and NOMA coexist, is considered. In such a network, four generic pairing
methods for NOMA with a heuristic pairing cost function are studied. When those methods cannot achieve a suitable performance level
OMA will be used for that subcarrier.
In contrast to \cite{new2}, here, we propose a joint resource allocation problem in which access technology selection, user pairing (for NOMA) or user selection (for OMA) and power allocation are jointly determined. We define a new set of optimization variables for access technology selection for each subcarrier. Also, the requirements of different services in 5G are very diverse and such diverse requirements must be considered in the resource allocation \cite{access}. Thus, our proposed utility function captures the tradeoff between data rate and the complexity of implementation of a given access technology.


The rest of this paper is organized as follows. First, Section~\ref{sec:model} describes the system model and problem formulation. Section~\ref{sec:raa} introduces the proposed resource allocation approach. Simulation results and analysis are presented in~\ref{sec:sr}. Finally, Section~\ref{sec:conclusion} concludes this paper.

\section{System Model and Problem Formulation}\label{sec:model}

Consider a cellular network with a single base station (BS) that services a set $\mathcal{K}$ of $K$ users in its own, specific coverage area. In this system, a set $\mathcal{S}$ of $S$ different SPs, provide service to their users sharing the BS. Each user belongs to one SP. Hence, we define $\mathcal{K}_s \subset \mathcal{K}$ as the subset of $K_s$ users subscribed to SP $s\in\mathcal{S}$. The purpose of introducing multiple SPs is to enable service customization as each SP has a minimum rate requirement for its own subscribed users.

The total available bandwidth is partitioned into a set $\mathcal{N}$ of $N$ subcarriers. We assume that the BS can switch between two access technologies (i.e., OMA and NOMA) for each subcarrier. $\beta_n$ represents the access technology selection parameter and is defined as follows:
 \begin{align}
  {{\beta }_{n}}=\left\{ \begin{array}{ll}
   0, &\text{if OMA is selected for subcarrier }\,n,  \\
   1, & \text{if NOMA is selected for subcarrier }\,n.
\end{array} \right.
\end{align}

Here, we consider that only two users can share a single subcarrier in NOMA \cite{csinoma}. Let $h_{k,n}$ be the wireless channel gain from BS to the user $k$ on subcarrier $n$, and $\sigma^2$ be the additive white Gaussian noise (AWGN) variance. Therefore, the downlink rate for a transmission over subcarrier $n\in \mathcal{N}$ will be given by:
\begin{align}\label{rate}
{R}_{n}(\bm{\alpha},\bm{\beta},\bm{p})=\sum\limits_{k\in \mathcal{K} }^{{}}{{\alpha }_{k,k,n}}\left[\log (1+\frac{{{p}_{k,n}}{{h}_{k,n}}}{{\sigma^2}})\right.+{{\beta }_{n}}\!\!\sum\limits_{\begin{smallmatrix}
   k_2\in \mathcal{K}  \\
   {{k}_{2}}\ne k  \\
\end{smallmatrix}}^{{}}\left.{\log (1+\frac{{{p}_{{{k}_{2}},n}}{{h}_{{{k}_{2}},n}}}{{{p}_{k,n}}{{h}_{{{k}_{2}},n}}+{\sigma^2}}){{\alpha }_{k,{{k}_{2}},n}}}\right],
\end{align}
where $p_{k,n}$ represents the power allocated by the BS to user $k$ over subcarrier $n$, and $\alpha_{k,k_2,n}$ is a binary variable given by:

\begin{eqnarray}
{{\alpha }_{k,k_2,n}}=\left\{ \begin{matrix}
   1, \lefteqn{~~\textrm{if }k \textrm{ and } k_2 \textrm{ are the first and second users\,of\,NOMA,}}\\&\!\!\!\textrm{ respectively, on subcarrier } n,~\forall k,k_2,~k\ne k_2,  \\
   1,  \lefteqn{~~\textrm{if }k_2\textrm{ is\,the\,first\,user\,of\,NOMA\,or\,user}}\\&\!\!\!\!\!\!\!\!\,\textrm{of OMA on subcarrier}\,n,~\forall k,k_2,~k=k_2,  \\
   0 \lefteqn{~~\textrm{otherwise.} } \\
\end{matrix} \right.\nonumber
\end{eqnarray}

In (\ref{rate}), the first term captures the rate of a user in OMA or the first user in NOMA, while the second term represents the rate of the second user that accounts for the interference of the first user in the NOMA case. When $\beta_n=0$, the second term is equal to zero and OMA is selected. The three-dimensional matrix $\bm{\alpha}=[\alpha_{k,k_2,n}]$, the vector $\bm{\beta}=[\beta_n]$, and the two-dimensional matrix $\bm{P}=[p_{k,n}]$ are optimization variables that capture the problem of resource allocation in our system.

Since additional processing is required at the receiver for successive interference cancellation when NOMA is selected, this additional complexity can be considered as a cost. Such cost would affect the access technology selection. Therefor, we propose a new utility function  to incorporate this cost for each subcarrier $n$, as follows:
\begin{align}
U_n(\bm{\alpha},\bm{\beta},\bm{P})=R_n(\bm{\alpha},\bm{\beta},\bm{P})-w\beta_n F_n(\bm{\alpha},\bm{\beta},\bm{P}),~\forall n\in \mathcal{N},\label{Un}
\end{align}
where $w$ is a normalizing factor to harmonize the cost and rate functions to the desired priority levels of these two parameters in designing the system and $F_n(\bm{\alpha},\bm{\beta},\bm{P})$ represents the total processing cost of NOMA.

For the case of NOMA, SIC receivers are required and this will incur an extra cost to the system as SIC receivers are computationally more complex. Depending on the structure of NOMA as well as some practical considerations, in order to define $F_n(\bm{\alpha},\bm{\beta},\bm{P})$, we focus on two major components that contribute to the NOMA complexity:
\begin{itemize}
\item \textit{SIC processing:}
Primarily, the complexity of SIC receivers is a function of the number of code layers in superposition coding scheme \cite{compnoma}. In our setup, we assume that a maximum of two users can share one subcarrier in NOMA, to keep the SIC complexity low, which is a common assumption in the NOMA literature, e.g., see \cite{twousern} and \cite{eenoma}. Consequently, two layers of superposition coding schemes will be used in this setup which leads to a constant cost in $F_n(\bm{\alpha},\bm{\beta},\bm{P})$  which is represented by parameter $A'$ hereinafter.
\item \textit{Error propagation:}
SIC receivers suffer from propagation errors. To alleviate such errors, more complex designs have been proposed such as multiple decision aided SIC receivers \cite{complexreci}. In \cite{complexreci}, it is shown that the required complexity is an increasing function of the requested bit error rate (BER) for SIC. On the other hand, hybrid automatic repeat request (HARQ) is necessary to have reliable communications when there is an error in signal detection. Compared to OMA, NOMA encounters higher HARQ probability and its HARQ design is more challenging \cite{harq} as it requires more processing, increases complexity, and imposes extra cost.
In traditional SIC receivers, the achievable BER corresponding to the detection of a strong
signal in presence of interference from a weaker signal, is inversely proportional to the experienced
SINR. Therefore, to capture these effects, our considered NOMA cost will also include a component that is an increasing function of the inverse SINR.
\end{itemize}
Therefore, we propose a NOMA cost function that properly captures these two components as follows:
\begin{align}\label{equobj2}
F_n(\bm{\alpha},\bm{\beta},\bm{P})=A'+\sum_{k\in\mathcal{K}}\sum_{k_2\in\mathcal{K},k_2\neq k}\left[\alpha_{k,k_2,n}\alpha_{k,k,n}G\left(\frac{h_{k,n}p_{k_2,n}+\sigma^2}{h_{k,n}p_{k,n}}\right)\right],
\end{align}
where $A'$ denotes the constant cost of considering two users per subcarrier in our setup and $G(.)$ is an increasing function of the inverse SINR. $G(\cdot)$ represents the additional complexity required to alleviate the error propagation to a desired level.
By increasing the difference between the power of the desired signal and the interference induced from other users over the same subcarrier, proper interference cancellation can be performed with lower complexity.
To simplify the optimization problem, we consider an increasing logarithm function for $G(.)$ as $G(\upsilon)=V'\log(\upsilon)$ where $V'$ is a positive scalar.
Considering (\ref{Un}) and (\ref{equobj2}), the final utility function is
\begin{align}\label{equobj}
\lefteqn{\!\!\!\!\!\!\!\!\!\!U_n(\bm{\alpha},\bm{\beta},\bm{P})=R_n(\bm{\alpha},\bm{\beta},\bm{P})-\beta_n\Big(A}\\&&+\sum_{k\in\mathcal{K}}\sum_{k_2\in\mathcal{K},k_2\neq k}\left[\alpha_{k,k_2,n}\alpha_{k,k,n}V\log\left(\frac{h_{k,n}p_{k_2,n}+\sigma^2}{h_{k,n}p_{k,n}}\right)\right]\Big) ,~\forall n\in \mathcal{N},\nonumber
\end{align}
where $V = V'w$ and $A = A'w$ hereinafter.
%

Considering virtualization at the radio access unit and assuming that multiple SPs share the BS, the isolation between SPs should be provided for resource allocation purposes \cite{MM}. This isolation requirement is represented as a rate requirement for each SP as follows \cite{ratecons},

\begin{align}\label{equc1}
\lefteqn{\sum\limits_{k\in {{\mathcal{K}}_{{{s}}}}}\sum\limits_{n\in \mathcal{N}}\Big[{ {{\alpha }_{k,k,n}}\log (1+\frac{{{p}_{k,n}}{{h}_{k,n}}}{{\sigma^2}}) }}
   \\&&\!\!\!\!+ \beta _{n}\!\!\sum\limits_{\begin{smallmatrix}
   {{k}_{1}}\in \mathcal{K}  \\
   {{k}_{1}}\ne k  \\
\end{smallmatrix}}^{{}}{\log (1+\frac{{{p}_{k,n}}{{h}_{k,n}}}{{{p}_{{{k_1}},n}}{{h}_{k,n}}+{\sigma^2}}){{\alpha }_{{{k_1}},k,n}{\alpha }_{k_1,k_1,n}}} \Big]\!\!>\!\!{{R}_{{{s}}}}, \forall s\in \mathcal{S},\nonumber
\end{align}
where $R_{s}$ is the minimum required rate for SP $s\in \mathcal{S}$. In (\ref{equc1}), the second term belongs to the case in which user $k$ is the second user of NOMA and shares the subcarrier $n$ with user $k_1$.

If subcarrier $n\in \mathcal{N}$ is selected for NOMA transmission, SIC should be applied at the receiver side. For performing SIC, the difference in power levels of NOMA users should be larger than a specific lower bound \cite{tolcons}. This implementation constraint can be written as follows
\begin{align}\label{equc2}
 {{\beta}_{n}}\sum\limits_{k\in \mathcal{K}}{\frac{{{h}_{k,n}}}{{\sigma^2}}{{\alpha }_{k,k,n}}\Big(\sum\limits_{\begin{smallmatrix}
 {{k}_{2}}\in \mathcal{K} \\
 {{k}_{2}}\ne k
\end{smallmatrix}}^{{}}{{{p}_{{{k}_{2}},n}}{{\alpha }_{k,{{k}_{2}},n}}} -{{p}_{k,n}} \Big)}>{{\beta}_{n}}{{P}_{\textrm{d}}}, \forall n \in \mathcal{N},
\end{align}
where $P_{\textrm{d}}$ represents the minimum required difference of the received power levels between two NOMA users.

Due to the fact that, in NOMA, the second user should have lower CSI compared to the first user, we should have $\alpha_{k,k_2,n}=0$ when $h_{k_2,n}>h_{k,n}$, which can be represented as
\begin{eqnarray}\label{equc3}
 \left(\frac{h_{{{k}_{2}},n}^{{}}}{h_{k,n}^{{}}}\right)\alpha _{k,{{k}_{2}},n}^{{}}\le 1,\quad \forall k,{{k}_{2}},k\neq k_2\in \mathcal{K},n\in \mathcal{N}.
\end{eqnarray}
Moreover, since only one user could be selected as the second user in NOMA, we need to have,
\begin{eqnarray}\label{equc4}
\sum\limits_{k\in {\mathcal{K}}}{\sum\limits_{\begin{smallmatrix}
 {{k}_{2}}\in \mathcal{K},{{k}_{2}}\ne k
\end{smallmatrix}}^{{}}{\alpha _{k,{{k}_{2}},n}^{{}}\le 1},\quad \forall n\in \mathcal{N}}.
\end{eqnarray}
When a given user is selected as the second user, the associated subcarrier should use NOMA. Moreover, only one user should be chosen as the second user, which imposes the following constraint
\begin{eqnarray}\label{equc5}
\beta _{n}^{{}}=\sum\limits_{k\in \mathcal{K}}{\sum\limits_{\begin{smallmatrix}
 {{k}_{2}}\in \mathcal{K},{{k}_{2}}\ne k
\end{smallmatrix}}{\alpha _{k,{{k}_{2}},n}\alpha _{k,{{k}},n},\quad \forall n\in \mathcal{N}}}.
\end{eqnarray}
Furthermore, the features of OMA and NOMA impose three additional constraints on the resource allocation problem.
First, each subcarrier is assigned to only one user in OMA and only one user should be selected as the first user in NOMA. Therefore, we have,
\begin{eqnarray}\label{equc6}
 \sum\limits_{k\in \mathcal{K}}{\alpha _{k,k,n}^{{}}\le 1,\quad \forall n\in \mathcal{N}}.
\end{eqnarray}
When user $k$ is not assigned to subcarrier $n\in \mathcal{N}$, its allocated power should be equal to zero. The mathematically expression of this practical consideration is
\begin{eqnarray}\label{equc7}
p_{k,n}^{{}}-\sum\limits_{{{k'}_{}}\in \mathcal{K}}^{{}}{\alpha _{k',{{k}_{}},n}^{{}}}{{P}_{\max }}\le 0,\quad \forall k\in \mathcal{K},n\in \mathcal{N}.
\end{eqnarray}
Moreover, the transmit power limitation of BS is controlled by
\begin{eqnarray}\label{equc8}
\sum\limits_{k\in \mathcal{K}}{\sum\limits_{n\in \mathcal{N}}{\alpha _{k,k,n}^{{}}(p_{k,n}^{{}}+\sum\limits_{\begin{smallmatrix}
 {{k}_{2}}\in \mathcal{K} \\
 {{k}_{2}}\ne k
\end{smallmatrix}}^{{}}{\alpha _{k,{{k}_{2}},n}^{{}}p_{{{k}_{2}},n}^{{}})}\le {{P}_{\max }}}}
\end{eqnarray}
Finally, by using (\ref{Un}), we can pose the following dynamic multiple access technology selection problem:

\begin{eqnarray}\label{equ11}
\underset{\bm{\beta} ,\bm{\alpha} ,\bm{P}}{\mathop{\max }}\sum\limits_{n\in \mathcal{N}}{{{R}_{n}}}-w\sum\limits_{n\in \mathcal{N}}{{{\beta}_{n}}}F_n \\
  \textrm{s.t.~} (\ref{equc1})-(\ref{equc8}),\nonumber
\end{eqnarray}
where $\beta _{n}^{{}}$ and $\alpha _{k,{{k}_{2}},n}^{{}}\in \{0,1\}$, and $p_{k,n}^{{}}\ge 0$. Problem \eqref{equ11} is a mixed integer assignment programming problem whose objective function is not concave and, hence, it is an NP-hard optimization problem. To solve this problem, next, we propose an iterative resource allocation algorithm.
\section{Proposed Algorithm and Performance Analysis}\label{sec:raa}
To solve (\ref{equ11}), we categorize the variables into two groups. The first group is discrete variables, including technology selection ($\beta_n,n\in\mathcal{N}$) and subcarrier allocation or user pairing for OMA or NOMA ($\alpha_{k,k_2,n},k,k_2\in\mathcal{K},n\in\mathcal{N}$) parameters. The second group of variables,  that includes the power levels $p_{k,n},k\in\mathcal{K},n\in\mathcal{N}$, is continuous. A two-step iterative algorithm is proposed to allocate the variables. In the first step, at iteration $t$, the values of $\alpha_{k,k_2,n}^{t}$ and $\beta_n^{t}$ are optimized considering the previous optimal value of $p_{k,n}^{t-1}$ at iteration $(t-1)$. In the second step, $p_{k,n}^{t}$ is optimized for a fixed value of $\alpha_{k,k_2,n}^{t}$ and $\beta_n^{t}$ obtained from the first step.
The mathematical expression of total iterative optimization procedure is as
\begin{align}\label{equ12}
{{\beta }^{0}},{{\alpha}^{0}}\to p^0\to {{\beta }^{1}},{{\alpha}^{1}},...,{{\beta }^{t}},{{\alpha}^{t}}\to {{p}^{t}}\to {{\beta }^{*}},{{\alpha }^{*}}\to {{p}^{*}}
\end{align}
This iteration continues until the algorithm converges and the following conditions are held
\begin{eqnarray}\label{equ13}
\left\| {{\beta }^{t}}-{{\beta }^{t-1}} \right\|\le {{\varepsilon }_{\beta }},\left\| {{\alpha }^{t}}-{{\alpha }^{t-1}} \right\|\le {{\varepsilon }_{\alpha }},\left\| {{p}^{t}}-{{p}^{t-1}} \right\|\le {{\varepsilon }_{p}},
\end{eqnarray}
where $0<{\varepsilon }_{\beta },{\varepsilon }_{\alpha },{\varepsilon }_{p}\ll 1$.
 \subsection{Technology Selection and Subcarrier Assignment Problem}\label{sec:first}
 The optimization problem at iteration $t$ for the fixed value of $p_{k,n}^{t-1}$ from the previous iteration is
 \begin{eqnarray}\label{equ14sub}
\lefteqn {\underset{\bm{\beta} ,\bm{\alpha} }
\max\sum\limits_{{n \in \mathcal{N}}}\sum\limits_{{k\in \mathcal{K}}}\alpha _{k,k,n}\Bigg(Y_1(p_{k,n}^{t-1})+\beta _{n}\sum\limits_{
 k\in \mathcal{K},{{k}_{2}}\ne k}Y_2(p_{k,n}^{t-1},p_{{{k}_{2}},n}^{t-1})\alpha _{k,{{k}_{2}},n}\Bigg)}\\&&-\sum\limits_{n\in \mathcal{N}}\beta _{n}\Big(A+V\sum_{k\in\mathcal{K}}\sum_{k_2\in\mathcal{K},k_2\neq k}\alpha_{k,k_2,n}\alpha_{k,k,n}\log(\frac{p_{k_2,n}^{t-1}h_{k,n}+\sigma^2}{h_{k,n}p_{k,n}^{t-1}})\Big),  \nonumber\\
 &&\textrm{s.t.} ~(\ref{equc1}),(\ref{equc3})-(\ref{equc6}),\nonumber
\end{eqnarray}
where $Y_1(p_{k,n}^{t-1})=\log (1+\frac{{p_{k,n}^{t-1}}{{h}_{k,n}}}{{\sigma^2}})$ and $Y_2(p_{k,n}^{t-1},p_{{{k}_{2}},n}^{t-1})=\log (1+\frac{{p_{{{k}_{2}},n}^{t-1}}{{h}_{{{k}_{2}},n}}}{{p_{k,n}^{t-1}}{{h}_{{{k}_{2}},n}}+{\sigma^2}})$ are the rate of users according to the optimal allocated power in the previous iteration.

To convert this problem into linear optimization, we use an auxiliary variable $u_{k,k_2,n}=\beta_n\alpha_{k,k,n}\alpha_{k,k_2,n}$. Note that, the defined variable $u_{k,k_2,n}$ depends on $\alpha_{k,k,n}$. As a result, a new constraint should be added in the optimization problem to prevent the unacceptable values for these dependent variables. In essence, $u_{k,k_2,n}$ cannot be equal to one when $\alpha_{k,k,n}=0$. Therefore, one of the key additional constraints is to have:
 \begin{eqnarray}\label{equ14n}
\alpha_{k,k,n}-u_{k,k_2,n}\geq 0.
\end{eqnarray}
Similarly, since $u_{k,k_2,n}$ cannot be equal to one when $\beta_n=0$, the following constraint should be satisfied:
 \begin{eqnarray}\label{equ14}
\beta_{n}-u_{k,k_2,n}\geq 0.
\end{eqnarray}

The optimization problem should be reformulated according to the new auxiliary variable.
For this purpose, by multiplying both sides of (\ref{equc5}) by $\beta_n$, we obtain:
\begin{eqnarray}\label{equ14bn}
\beta_n^2=\sum_{k\in\mathcal{K}}\sum_{k_2\in\mathcal{K}}u_{k,k_2,n}.
\end{eqnarray}
Since the variable $\beta_n$ is binary, we have $\beta_n^2=\beta_n$. Therefore, (\ref{equ14bn}) is used in (\ref{equ14sub}).
Also, $u_{k,k_2,n}$ is equal to zero when $\alpha_{k,k_2,n}$ is zero. Thus, $\alpha_{k,k_2,n}$ in (\ref{equc3}) and (\ref{equc4}) can be replaced by $u_{k,k_2,n}$.
Finally, (\ref{equ14sub}) is transformed into
%

 \begin{align}\label{equ14lin}
\lefteqn {\max \underset{\bm{\beta},\bm{u} ,\bm{\alpha} }\ \sum\limits_{{n \in \mathcal{N}}}\sum\limits_{{k\in \mathcal{K}}}\!\!\alpha _{k,k,n}^{{}}~\!\!Y_1(p_{k,n}^{t-1})+^{{}} \sum\limits_{\begin{smallmatrix}
 k\in \mathcal{K} \\
 {{k}_{2}}\ne k
 \\
\end{smallmatrix}}Y_2(p_{k,n}^{t-1},p_{{{k}_{2}},n}^{t-1})u _{k,{{k}_{2}},n}}\\&-\sum\limits_{n\in \mathcal{N}}\Bigg( \beta_n A+V\sum_{k\in\mathcal{K}}\sum_{k_2\in\mathcal{K},k_2\neq k}u_{k,k_2,n}\log\bigg(\frac{h_{k,n}p_{k_2,n}^{t-1}+\sigma^2}{h_{k,n}p_{k,n}^{t-1}}\bigg)\Bigg),\nonumber
\nonumber\\&
 \textrm{s.t.}~ \alpha_{k,k,n}-u_{k,k_2,n}\geq 0\quad\forall k,{{k}_{2}}\in \mathcal{K},n\in \mathcal{N},
\nonumber\\&
 \beta_{n}-u_{k,k_2,n}\geq 0\quad\forall k,{{k}_{2}}\in \mathcal{K},n\in \mathcal{N},
\nonumber\\&
 \sum\limits_{{n \in \mathcal{N}}}\sum\limits_{{k\in \mathcal{K}_{s}}} \alpha _{k,k,n}^{{}}~ Y_1(p_{k,n}^{t-1})+^{{}}\sum\limits_{\begin{smallmatrix}
 k_1\in \mathcal{K} \\
 {{k}_{1}}\ne k
 \\
\end{smallmatrix}}Y_2(p_{k_1,n}^{t-1},p_{{{k}},n}^{t-1})u _{k_{1},{{k}},n}^{{}}>R_s, \forall s\in \mathcal{S},
\nonumber\\&
  \left(\frac{h_{{{k}_{2}},n}^{{}}}{h_{k,n}^{{}}}\right)u _{k,{{k}_{2}},n}^{{}}\le 1,\quad \forall k,{{k}_{2}}\in \mathcal{K},n\in \mathcal{N},
 \nonumber\\&
 \sum\limits_{k\in {\mathcal{K}}}{\sum\limits_{\begin{smallmatrix}
 {{k}_{2}}\in \mathcal{K},{{k}_{2}}\ne k
\end{smallmatrix}}^{{}}{u _{k,{{k}_{2}},n}^{{}}\le 1},\quad \forall n\in \mathcal{N}},
\nonumber\\&
 \beta_n=\sum_{k\in\mathcal{K}}\sum_{k_2\in\mathcal{K},k_2\neq k}u_{k,k_2,n},
\nonumber\\&
\sum\limits_{k\in \mathcal{K}}{\alpha _{k,k,n}^{{}}\le 1,\quad \forall n\in \mathcal{N}}.\nonumber
\end{align}

This problem is a linear integer programming. There are different approaches to solve linear optimization, among them, the interior-point method has gained much more attention due to its simplicity. Here the linear integer programming problem of (\ref{equ14lin}) is solved using CVX \cite{cvxlin}, which uses the interior-point method.


\subsection{Power Allocation Problem}\label{sec:pa}
In Step 2, given to the best access selection and subcarrier allocation for users derived in Step 1 ($\beta_{n}^t$ and $\alpha_{k,k,n}^t$, $u_{k,k_2,n}^t$), the BS should decide on its power allocation across subcarriers. In fact, by substituting the derived values for the optimization variables in (\ref{equ11}) and omitting the constraints which do not depend on the power allocation, the optimization problem for power allocation becomes
\begin{align}\label{equpower}
\max_{\bm{P}}&\sum\limits_{n\in \mathcal{N}}\sum\limits_{k\in \mathcal{K} }{\alpha_{k,k,n}^t}\log (1+\frac{{p_{k,n}}{{h}_{k,n}}}{{\sigma^2}})\\
+&\sum\limits_{n\in \mathcal{N}}\sum\limits_{k\in \mathcal{K} }\sum\limits_{\begin{smallmatrix}
	k\in \mathcal{K}  \\
	{{k}_{2}}\ne k  \\
	\end{smallmatrix}}^{{}}u_{k,k_2,n}^t\big(\log ({p_{k,n}}{{h}_{{{k}_{2}},n}}+{\sigma^2}+{p_{{{k}_{2}},n}}{{h}_{{{k}_{2}},n}})-\log ({p_{k,n}}{{h}_{{{k}_{2}},n}}+{\sigma^2})\big)\nonumber \\
-&V\sum_{n\in\mathcal{N}}\sum_{k\in\mathcal{K}}\sum_{\begin{smallmatrix}
	k\in \mathcal{K}  \\
	{{k}_{2}}\ne k  \\
	\end{smallmatrix}}^{{}} u_{k,k_2,n}^t\big(\log(p_{k_2,n}h_{k,n}+N_o)-\log(p_{k,n}h_{k,n})\big)\nonumber\\
& \textrm{s.t.}~ \ (\ref{equc1}),(\ref{equc2}),(\ref{equc7}),(\ref{equc8}) \nonumber
\end{align}
where $\bm{P}$ is a $K\times N$ matrix in which each element at row $k$ and column $n$ is equal to $p_{k,n}$.

By applying the DC algorithm \cite{dc}, we can approximate the negative logarithmic terms in the objective function with affine functions, as follows

\begin{eqnarray}\label{equpower}
\lefteqn
  {\!\!\!\!\!J(\bm{P})=\sum\limits_{n\in \mathcal{N}}\sum\limits_{k\in \mathcal{K} }{\alpha_{k,k,n}^t}\log (1+\frac{{p_{k,n}}{{h}_{k,n}}}{{\sigma^2}})  +{}\sum\limits_{\begin{smallmatrix}
   k\in \mathcal{K}  \\
   {{k}_{2}}\ne k  \\
\end{smallmatrix}}^{{}}{u _{k,{{k}_{2}},n}}\Big(\log ({p_{k,n}}{{h}_{{{k}_{2}},n}}+{\sigma^2}+{p_{{{k}_{2}},n}}{{h}_{{{k}_{2}},n}})}\nonumber\\&&-\big(\log(p_{k,n}^{t_2-1}h_{k2,n}+\sigma^2)+\frac{(p_{k,n}-p_{k,n}^{t_2-1})h_{k_2,n}}{p_{k,n}^{t_2-1}h_{k_2,n}+\sigma^2}\big)\Big)+V\sum_{n\in\mathcal{N}}\sum_{k\in\mathcal{K}}\sum_{k_2\in\mathcal{K},k_2\neq k}\!\!\!\!\!\!u_{k,k_2,n}\Big(-\big(\log(p_{k_2,n}^{t_2-1}h_{k,n}+\sigma^2)\nonumber\\&&+(p_{k_2,n}-p_{k_2,n}^{t_2-1})\frac{h_{k,n}}{p_{k_2,n}^{t_2-1}h_{k,n}+\sigma^2}\big)+\log(p_{k,n}h_{k,n})\Big).
\end{eqnarray}
The same approach is used for constraint (\ref{equc1})
\begin{eqnarray}\label{equc1'}
&
\sum\limits_{n\in \mathcal{N}}\sum\limits_{k\in {{\mathcal{K}}_{{{S}}}}}\Biggr({{ {{\alpha }_{k,k,n}^t}\log (1+\frac{{{p}_{k,n}}{{h}_{k,n}}}{{\sigma^2}}) }}
   + \sum\limits_{\begin{smallmatrix}
   {{k}_{1}}\in \mathcal{K}  \\
   {{k}_{1}}\ne k  \\
\end{smallmatrix}}{{{u }_{{{k_1}},k,n}}}\Big({\log (p_{k_1,n}h_{k,n}+\sigma^2+p_{k,n}h_{k,n})}
\nonumber\\&
-(\log(p_{k_1,n}^{t_2-1}+\sigma^2)+\frac{(p_{k_1,n}-p_{k_1,n}^{t_2-1})h_{k,n}}{p_{k_1,n}^{t_2-1}h_{k,n}+\sigma^2})\Big)
\Biggr)>{{R}_{{{s}}}},~\forall s\in \mathcal{S}.
\end{eqnarray}

Finally, the transformed power allocation optimization problem to the convex one, is
\begin{align}\label{equpowerprob}
\lefteqn{~~~~~~~~~~~~~~~~~\max_{\bm{P}}J({\bm{P}})}\\
&&\!\!\!\!\!\!\!\!\!\!\!\!\!\!\!\!\!\!\!\!\!\!\!\!\!\!\!\!\!\!\!\! \textrm{s.t.~} \ (\ref{equc1'}),(\ref{equc2}),(\ref{equc7}),(\ref{equc8}). \nonumber
\end{align}

The Lagrangian multiplier method is applied to the convex problem (\ref{equpowerprob}) and, then, the Lagrange is:

\begin{eqnarray}
\lefteqn{ L(\bm{P},\bm{\lambda,\gamma,\zeta,\eta})=\sum_{k}\sum_{n}\alpha_{k,k,n}^t\log(1+\frac{p_{k,n}h_{k,n}}{\sigma^2})+\sum_{k_2,k_2\neq k}u_{k,k_2,n}^t\Big(\log(p_{k,n}h_{k_2,n}+p_{k_2,n}h_{k_2,n}+\sigma^2)}\nonumber\\&&
-\frac{h_{k_2,n}}{p_{k,n}^{t_2-1}h_{k_2,n}+\sigma^2}(p_{k,n}-p_{k,n}^{t_2-1})-\log(p_{k,n}^{t_2-1}h_{k_2,n}+\sigma^2)\Big)
\nonumber\\&&+V\sum_{n\in\mathcal{N}}\sum_{k\in\mathcal{K}}\sum_{k_2\in\mathcal{K},k_2\neq k}u_{k,k_2,n}^t\Big(-\big(\log(p_{k_2,n}^{t_2-1}h_{k,n}+\sigma^2)+\frac{(p_{k_2,n}-p_{k_2,n}^{t_2-1})h_{k,n}}{p_{k_2,n}^{t_2-1}h_{k,n}+\sigma^2}\big)+\log(p_{k,n}h_{k,n})\Big)\nonumber\\&&
+\sum_{s\in \mathcal{S}}\lambda_{s}\Biggr(\sum_{k \in \mathcal{K\textrm{s}}}\sum_n\alpha_{k,k,n}^t\log(1+\frac{p_{k,n}h_{k,n}}{\sigma^2})+\sum_{k_1,k_1\neq k}u_{k_1,k,n}^t\Big(\log(p_{k,n}h_{k,n}+p_{k_1,n}h_{k,n}+\sigma^2)\nonumber\\&&
-\log(p_{k_1,n}^{t_2-1}h_{k,n}+\sigma^2)-\frac{h_{k,n}}{p_{k_1,n}^{t_2-1}h_{k,n}}(p_{k_1,n}-p_{k_1,n}^{t_2-1})\Big)-R_s\Biggr)
\nonumber\\&&+\sum_{n}\gamma_n\left(\sum_{k}\frac{h_{k,n}}{\sigma^2}\Big(\sum_{k_2,k_2\neq k}p_{k_2,n}u_{k,k_2,n}^t-\beta_n^t p_{k,n}\alpha_{k,k,n}^t\Big)-\beta_n^tp_\textrm{d}\right)\nonumber
\end{eqnarray}
\begin{eqnarray}\label{equpowerlag}
-\sum_{k_2}\sum_{n}\zeta_{k_2,n}(p_{k_2,n}-\sum_k u_{k,k_2,n}^tP_{\textrm{max}})-\eta\Big(\sum_{k}\sum_{n}\alpha_{k,k,n}^tp_{k,n}+\sum_{k_2,k_2\neq k}u_{k,k_2,n}^tp_{k_2,n}-P_{\textrm{max}}\Big).\nonumber\\
\end{eqnarray}

In this method, the primal (maximizing $L(\bm{P})$ to find $\bm{P}$) and dual (minimizing $L(\bm{\lambda,\gamma,\zeta,\eta})$ to find Lagrange multipliers) problems are solved iteratively until changes in the variables are negligible and the iterative algorithm converges to a fixed point. Since the optimization problem (\ref{equpowerprob}) is convex, this fixed point will be the optimal.

Next, in Proposition 1, for fixed access technology selections, we derive closed-form expressions for the optimal power allocation strategy as a function of the NOMA processing cost coefficient ($V$), channel gains, and Lagrange multipliers. The Lagrangian multipliers (dual variables) are obtained by solving the dual problem using the gradient method.

\begin{prop}%
Given the access technology selection for subcarrier $n$, the optimal power allocation will be given by:

\begin{itemize}
 \item If $\beta_n=0$ and $\alpha_{k,k,n}=1$,
 \begin{align}\label{equpoweroma}
p_{k,n}=\left[-\frac{1+\lambda_{s'}}{D_{k,n}}-\frac{\sigma^2}{h_{k,n}}\right]^+
\end{align}
where $D_{k,n}=-\zeta_{k,n}-\eta$.

\item If $\beta_n=1$, $\alpha_{k_1,k_1,n}=1$, and $\alpha_{k_1,k_2,n}=1$,
 \begin{align}\label{equpowernoma}
p_n=\left[-\frac{1+\lambda_{s'}}{D_{k_2,n}}-\frac{\sigma^2}{h_{k_2,n}}\right]^+
 \end{align}
 and
  \begin{align}\label{equpowernoma2}
p_{k_1,n}=\frac{-({{h}_{k_1,n}}(1+{{\lambda }_{s'}}+V)+{\sigma^2}Q)\pm \sqrt{{{({{h}_{k_1,n}}(1+{{\lambda }_{s'}}+V)+{\sigma^2}Q)}^{2}}-4V{\sigma^2}Q{{h}_{k_1,n}}}}{2Q{{h}_{k_1,n}}}
\end{align}
where
\begin{align}
&p_n=p_{k_1,n}+p_{k_2,n}, \forall k_1\in\mathcal{K}_{s'}, \forall k_2\in\mathcal{K}_{s''},\\ &Q=\frac{-{{D}_{k_2,n}}(1+{{\lambda }_{s''}})}{1+{{\lambda }_{s'}}}+{{D}_{k_1,n}},\nonumber\\ &D_{k_1,n}=-(1+\lambda_{s''})\frac{-h_{k_2,n}}{p_{k_1,n}^{t_2-1}h_{k_2,n}+\sigma^2}+\gamma_n\frac{h_{k_1,n}}{\sigma^2}-\zeta_{k_1,n}-\eta,~\textrm{and}\nonumber\\  &D_{k_2,n}=-V\frac{h_{k_2,n}}{p_{k_2,n}^{t_2-1}h_{k_2,n}+\sigma^2}+\gamma_n\frac{h_{k_1,n}}{\sigma^2}-\zeta_{k_2,n}-\eta.\nonumber
\end{align}
\end{itemize}
\begin{proof}
See Appendix \ref{appen}.
\end{proof}
\end{prop}

According to (\ref{equpoweroma}), the power allocation of OMA users is based on a pseudo water-filling algorithm. Similarly, in (\ref{equpowernoma}), the power allocation is in a water-filling manner. However, the channel gain of the second user in NOMA plays a key role in the total power allocated to both NOMA users in one subcarrier.
	
To study the effects of $V$ on the power allocation of OMA and NOMA users, we assume Lagrange  multipliers to be fixed. By increasing $V$, the absolute value of $D_{k_2,n}$ increases which leads to a lower value for the first term in (\ref{equpowernoma}). As a result, the total power allocated to the NOMA users decreases. Considering a fixed total power $P_{\textrm{max}}$ for the BS, decreasing the total power allocated to the NOMA users leads to a higher allocated power to the OMA users. In other words, Proposition 1 shows that the optimal power allocated to OMA  increases (thus the allocated power to NOMA decreases) when the cost of NOMA grows.

Two other important aspects to evaluate the performance of the proposed resource allocation are its convergence and computational complexity. Regarding the convergence, the proposed iterative algorithm uses the block coordinate descent (BCD) method in which one group of variables is optimized and the others are assumed to be fixed. In \cite{razavi}, it is shown that the convergence of BCD is guaranteed when the variable groups are updated by a successive sequence of approximations of the objective function like strictly convex local approximations. Therefore, applying successive convex approximation, the convergence of algorithm is guaranteed. However, the convergence is guaranteed to a local optimum, which may not be the global optimum.

Regarding the computational complexity, in a primal-dual interior point method for linear programming of the first step of the proposed algorithm, $O(\sqrt{n_v}n_s)$ (where $n_v$ is the number of variables and $n_s$ is the size of the problem data) iterations are required in the worst case to obtain a solution that can be transformed easily into an optimal basic feasible solution \cite{LP}. The major computation in each iteration of the primal-dual interior point method is the construction and Cholesky factorization of a symmetric and positive definite matrix of size $m$ by $m$, where $m$ is the number of linear equality constraints. The computational complexity of Cholesky factorization is $O(m^3)$ in the worst case \cite{complex}. The complexity of the Lagrangian method for power allocation is $O(n_v/\epsilon_{\textrm{sub}}^2)$ where $1/\epsilon_{\textrm{sub}}^2$ is the number of iterations of sub-gradient method to find a $\epsilon_{\textrm{sub}}$-suboptimal point. On the other hand, the convergence of DC method is achieved by complexity of $O(\log(1/\epsilon_{\textrm{dc}}))$ where $\epsilon$ is the stopping criterion \cite{dccompl}. Consequently, the total complexity of power allocation algorithm is $O(n_v/(\epsilon_{\textrm{sub}}^2\epsilon_{\textrm{dc}}))$. This proves that the complexity of our proposed scheme only grows polynomially with the number of variables, which is a considerable improvement over direct search methods with exponential complexities.

\section{Simulation Results}\label{sec:sr}

For our simulations, we consider a network with 10 subcarriers and 20 users. Two SPs seek to provide service to their subscribed users. Users are uniformly distributed in a square area with a unit length. The channel gains of users are modeled by assuming large and small scale fading as $h_{k,n}=d_k^{-\alpha}s_{k,n}$, where $d_k$ is the distance between the BS and user $k$, $\alpha=3$ and $s_{k,n}$ has an exponential distribution with unit variance in Rayleigh fading channel. In our simulation setup, we consider normalized noise, i.e., $\sigma^2=1$, and $P_{\textrm{max}}=100$ W, and the required minimum power difference for SIC is set equal to $P_{\textrm{d}}=0.01$. The cost of NOMA processing is set to $A=V=2$. In our simulation results, when the constraints of the optimization problem cannot be satisfied simultaneously for a given CSI, the utility is set to zero. All statistical results are averaged over a large number of independent runs.
\begin{figure}\label{fig:2}
\centering
    \subfigure[] []{{\includegraphics[width=.7\textwidth,clip=true,trim=30 180 50 180]{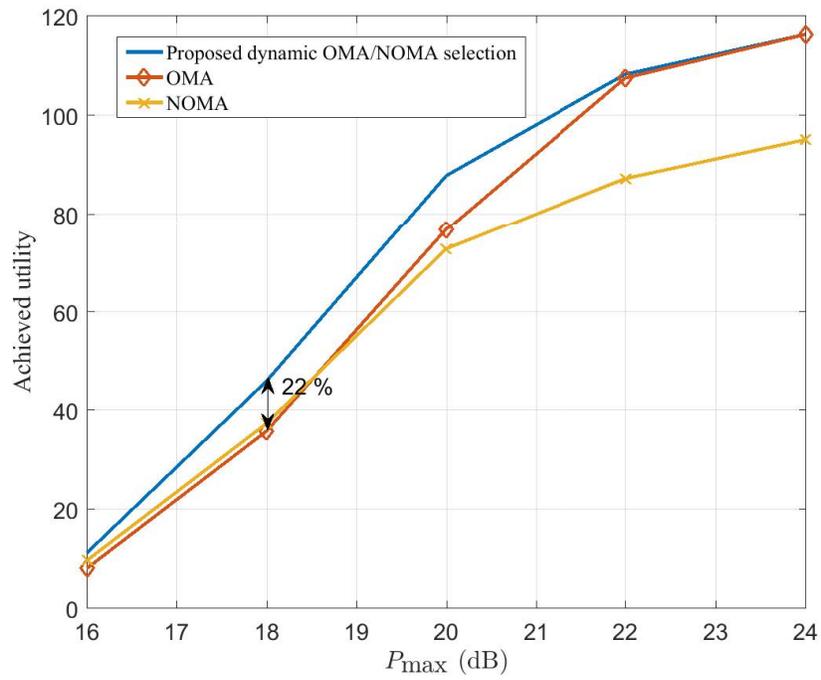}\label{fig:2a} }}\\
    \subfigure[] []{{\includegraphics[width=.7\textwidth,clip=true,trim=30 180 50 180]{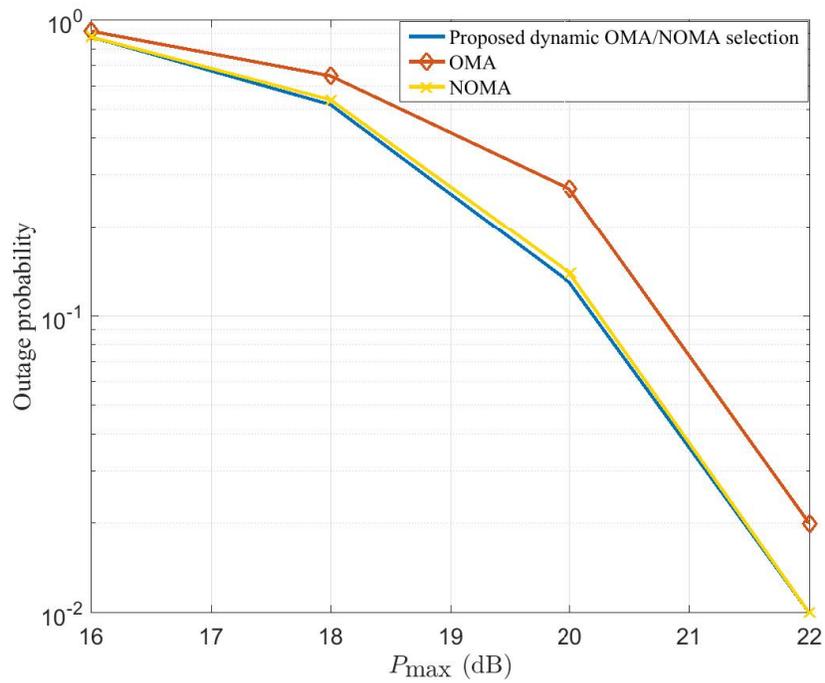} \label{fig:2b} }}
\caption{(a) Utility performance and (b) Outage probability of different access technologies versus different values for maximum transmit power.}
\end{figure}
To show the benefits of dynamic multiple access technology selection, the achieved utility of the proposed algorithm is compared to the NOMA and OMA cases. The optimization of OMA and NOMA, respectively, can be obtained by setting $\beta_n=0$ and $\beta_n=1$ in (\ref{equ14lin}) and (\ref{equpowerprob}).

In Fig. \ref{fig:2a}, the value of the proposed objective function which is a difference between the rate and the cost of NOMA processing is shown for different values of maximum transmit power where $R_s=48$ bps/Hz. As can be seen from Fig. \ref{fig:2a}, our proposed scheme outperforms the OMA and NOMA technologies. NOMA can support more than one user in each subcarrier and thus can facilitate meeting the QoS requirements of SPs compared to OMA. Therefore, it can be observed that NOMA enhances the performance for low power ranges despite of its processing cost. Note that in $18$ dB point where the feasibility of (\ref{equc1}) is very sensitive to the resource allocation strategy due to the dominant constraint (\ref{equc8}), the performance of our proposed scheme is approximately $22\%$ higher than NOMA.

We define the outage probability in (\ref{equc1}) as the probability that the rate of at least one SP is lower than its minimum required rate. This performance metric is studied in Fig.~\ref{fig:2b}. As expected, our proposed scheme offers the lowest outage probability due to the flexibility in access technology selection.
Based on Fig.~\ref{fig:2b}, for $R_s=48$ bps/Hz, $P_{\textrm{max}}=16$ dB is not enough to satisfy the rate requirements for SPs and the optimization problem will be mostly infeasible which explains the very high outage at those values. As can be seen for $P_{\textrm{max}}=20$ dB, our proposed strategy achieves approximately half of the outage probability of NOMA and OMA. The flexibility offered by our proposed scheme to select between OMA and NOMA leads to the higher probability to satisfy the minimum rate of SPs, and, thus, lower outage probability compared to the OMA and NOMA.

\begin{figure}\label{fig:1}
\centering
    \subfigure[] []{{\includegraphics[width=.7\textwidth,clip=true,trim=30 160 50 180]{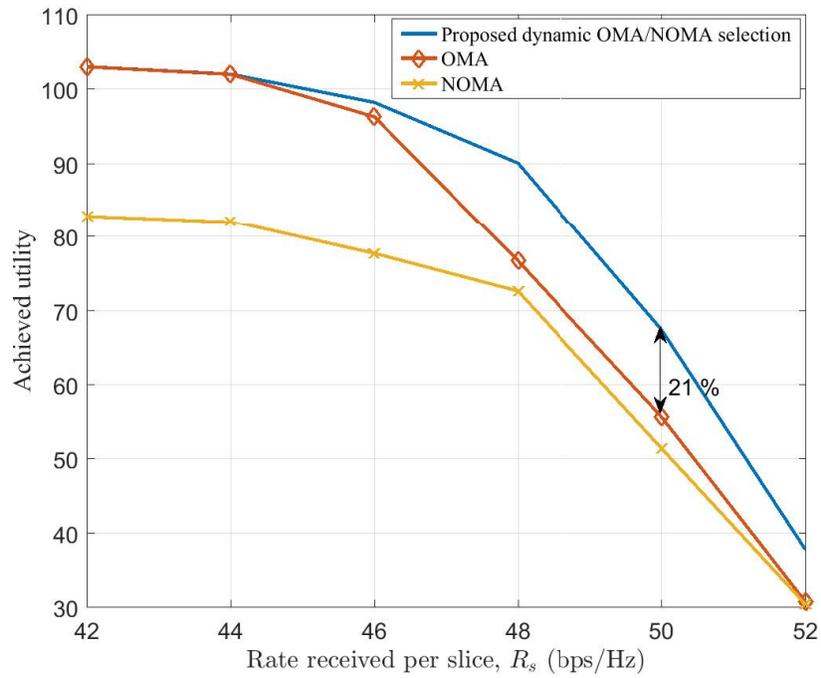}\label{fig:1a} }}\\
    \subfigure[] []{{\includegraphics[width=.7\textwidth,clip=true,trim=30 160 50 180]{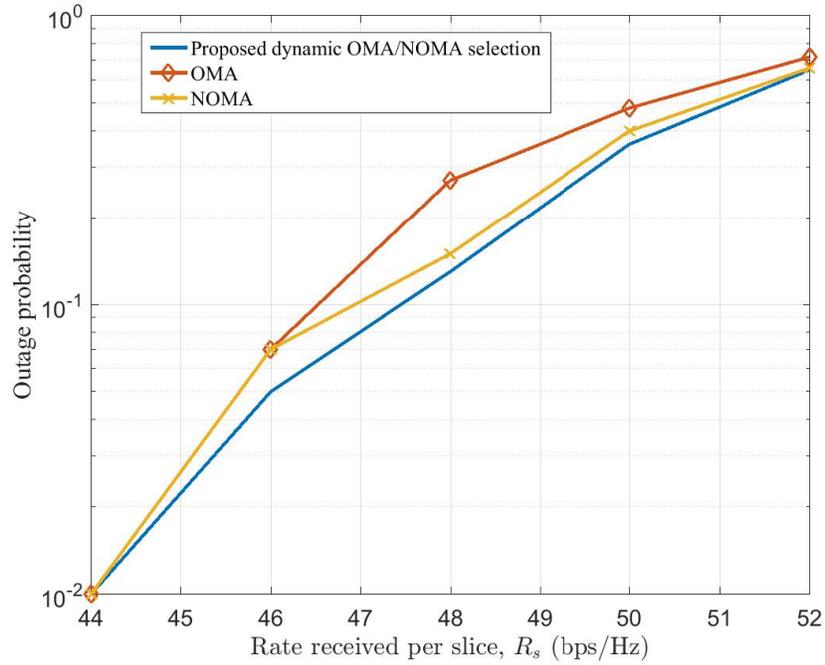} \label{fig:1b} }}
\caption{(a) Utility performance and (b) Outage probability of different access technologies versus different values for QoS of SPs.}
\end{figure}

 Fig. \ref{fig:1a} and Fig. \ref{fig:1b} analyze the impact of the SPs' QoS requirement parameter $R_s$ on the achieved utility and outage probability. As expected, the proposed algorithm improves the performance in terms of both utility function and outage probability compared to the pure OMA or NOMA cases. As can be seen in Fig. \ref{fig:1a}, for low ranges of $R_s$, the proposed scheme will mostly choose OMA since the probability of infeasibility is very low. In these cases, OMA is the best choice since the cost is lower.
 By increasing the minimum rate requirement, the feasibility region is shrunk in any access technology and outage probability increases. Meanwhile, in OMA the outage probability increases more quickly compared to the NOMA case since there are more restrictions as each subcarrier can be allocated to only one user. In other words, the ability of NOMA to allocate two users to each subcarrier results in a higher opportunity to satisfy the minimum required rate of SPs. Compared to the pure OMA and NOMA cases, our proposed strategy for resource allocation achieves lower outage probability since dynamic access technology selection expands the feasibility region. Also, the ability of the proposed scheme to choose between OMA with no processing cost as well as NOMA leads to performance improvement in the terms of defined utility compared to the pure NOMA technology. For example, when the feasibility of (\ref{equc1}) is the dominant constraint (e.g. for $R_s=50$), the proposed strategy for dynamic access technology selection yields utility gains of approximately $21\%$ compared to OMA.

\begin{figure}
\centering
  \includegraphics[width=.7\linewidth,clip=true,trim=30 160 50 180]{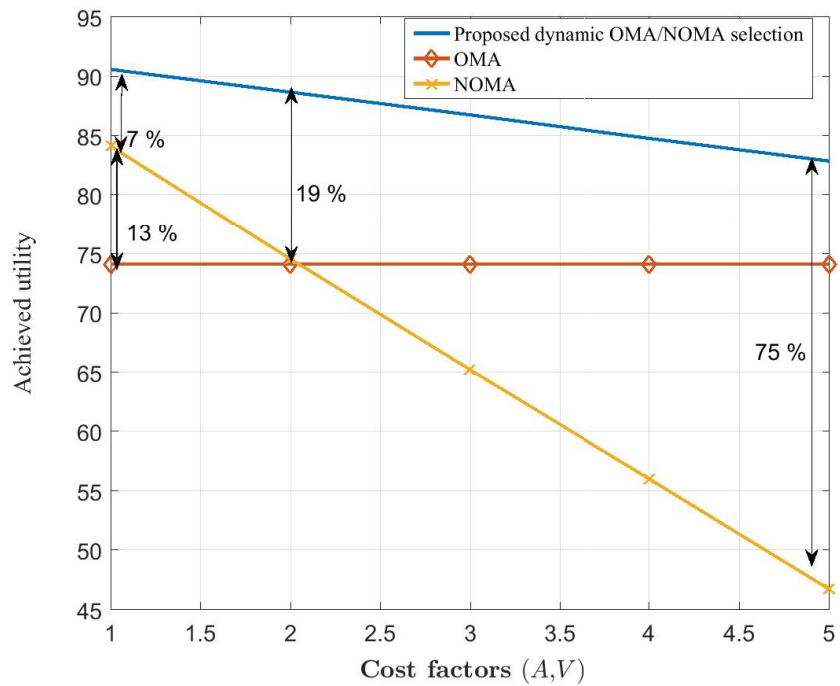}
  \caption {Utility performance of different access technologies versus different values for $A$ and $V$ for $R_s=48$.}
  \label{fig:5}
\end{figure}

\begin{figure}
\centering
  \includegraphics[width=.7\linewidth,clip=true,trim=30 160 50 180]{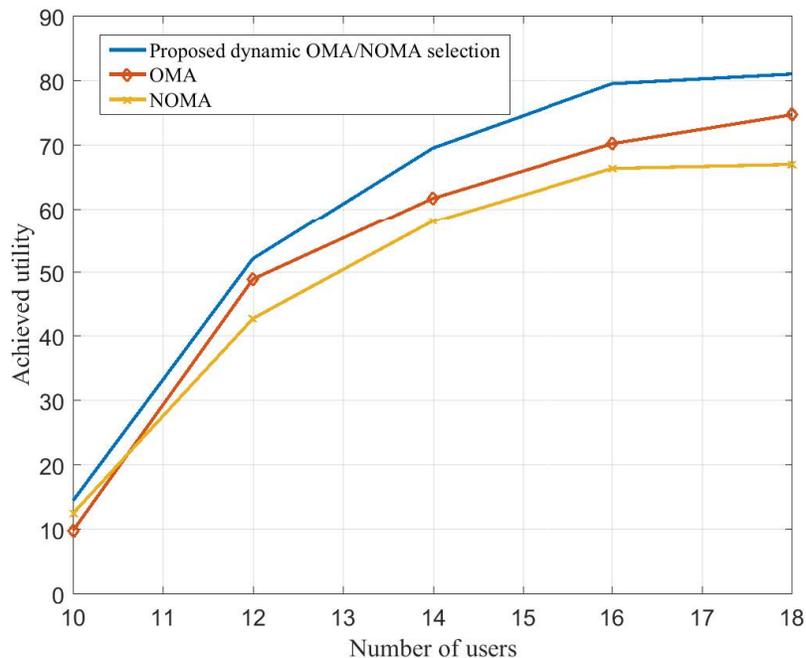}
  \caption {Utility performance of different access technologies versus number of users.}
  \label{fig:densityuser}
\end{figure}

%
%
%

\begin{figure}
\centering
  \includegraphics[width=.7\linewidth,clip=true,trim=30 160 50 180]{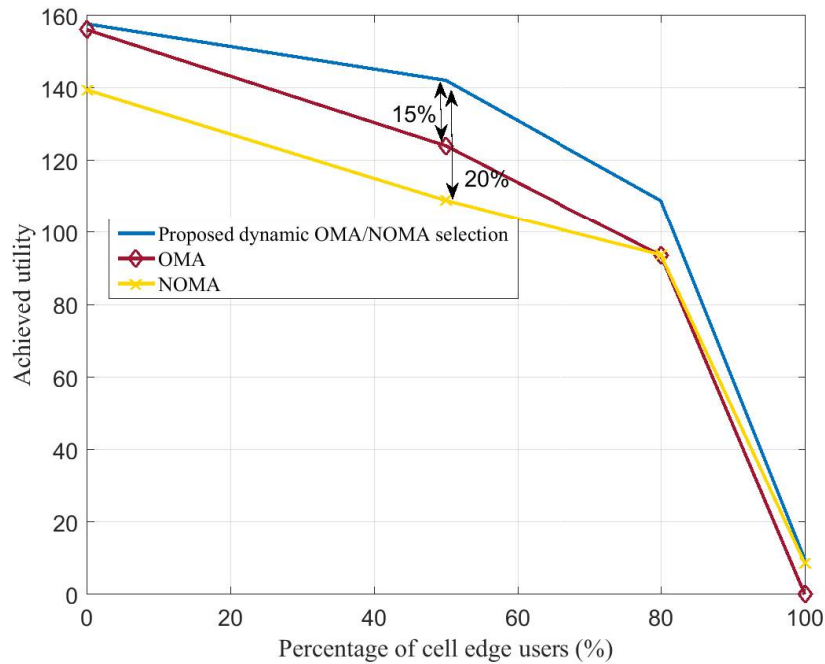}
  \caption {Utility of different access technologies versus percentage of cell edge users.}
  \label{fig:densityuti}
\end{figure}
\begin{figure}
\centering
  \includegraphics[width=.7\linewidth,clip=true,trim=30 160 50 180]{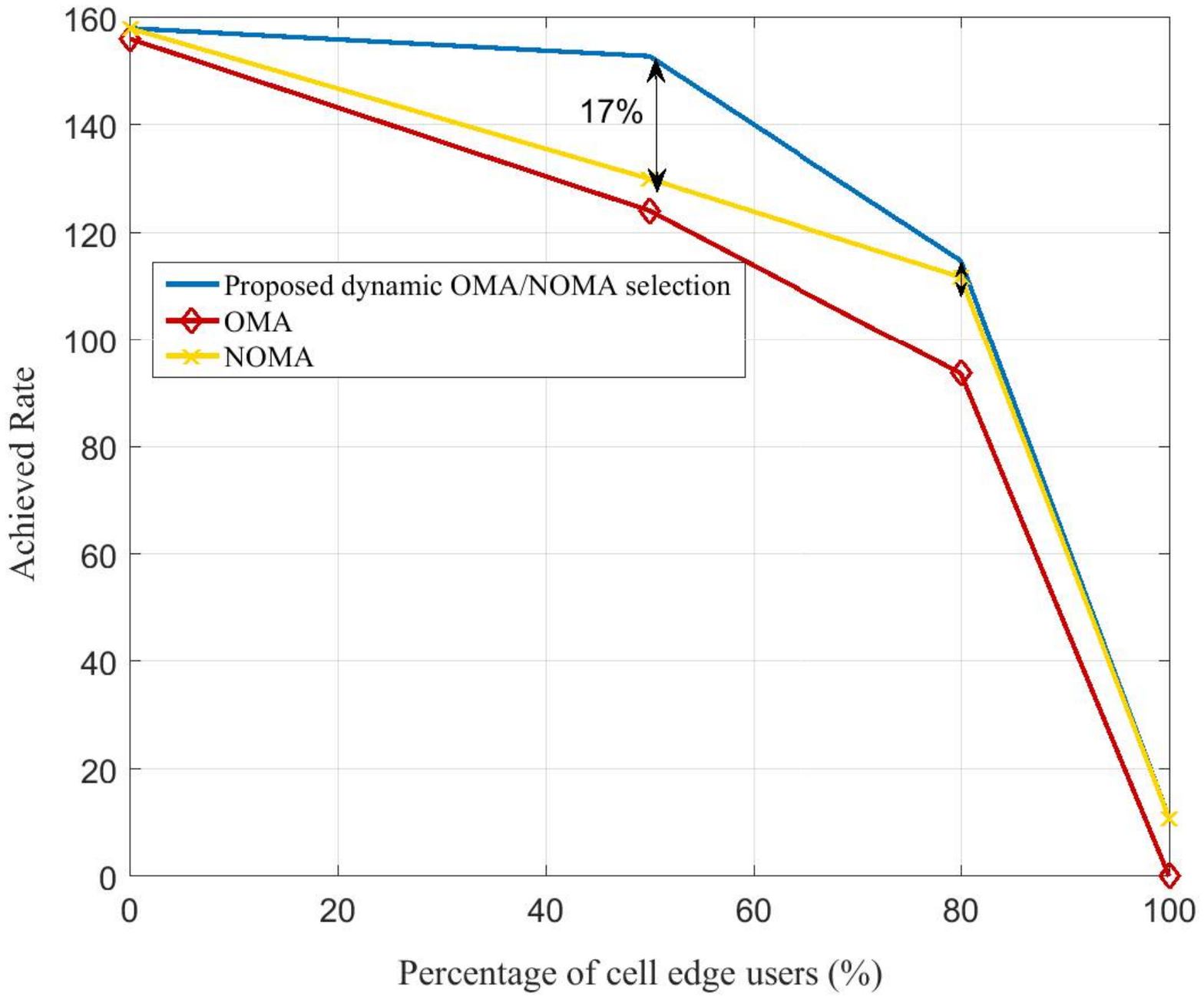}
  \caption {Rate performance of different access technologies versus percentage of cell edge users.}
  \label{fig:densityrate}
\end{figure}

In Fig. \ref{fig:5}, we evaluate the effects of the processing cost values on the system performance by presenting the value of the utility function for different values of $A$ and $V$. From Fig. \ref{fig:5}, by increasing the value of these parameters, the proposed scheme can enhance the performance up to $75\%$ compared to the NOMA scenario.
For low values of $A$ and $V$, NOMA achieves up to $13\%$ utility improvement compared to the OMA case. In fact, when taking into account both the rate and processing cost, on their own, neither NOMA nor OMA will be suitable for all use cases. As can be seen, our proposed scheme with flexible access technology improves the utility from $7\%$ to $19\%$.

%

In Fig. \ref{fig:densityuser}, we study effect of the number of users on the performance of proposed method. \ref{fig:densityuser}. In this figure, the other parameters are similar to those used in Fig. \ref{fig:2a} where the maximum power is set to the $100$ W. In general, by increasing the number of users in the system, the utility of all schemes increases due to the users' diversity gain. When $N<11$, NOMA has higher utility compared to OMA despite the processing cost. This is because NOMA can expand the feasibility region as two users can be supported in each subcarrier. For $N\geq11$, the performance of OMA improves owing to the users' diversity gain and up to $10\%$ higher performance is achieved compared to NOMA. This means the processing cost for NOMA influences its effectiveness compared to OMA. However, our proposed dynamic scheme outperforms both the NOMA and OMA cases for any number of users in the network.

In Fig. \ref{fig:densityuti} and Fig. \ref{fig:densityrate}, we study the effect of distribution density of users in the coverage region of BS on the achievable utility
 and total rate, respectively.
 For this purpose, we divide the coverage region into central and edge regions. The central region is the square with the half length of total coverage area, which is located at the center of the coverage region, and the rest is the edge region. The users in the edge region are called cell-edge users. In these figures, the x-axis represents the percentage of users in the cell-edge region. Other parameters are set as in Fig. \ref{fig:2a} where $P_{\textrm{max}}=100$ W.
 In Fig. \ref{fig:densityuti}, when no users are located in the cell-edge region, NOMA achieves the lowest performance among the three schemes in the terms of achieved utility, in presence of NOMA cost.
Furthermore, it is shown that, the utility performance of OMA is equal to the utility performance of NOMA when $80\%$ of users are distributed at the cell edge, even though NOMA imposes an extra processing cost. As a result, we can conclude that the performance of OMA is severely degraded when all users are at the cell edge (point $100\%$). However, NOMA can provide a better performance since it can support more users, and consequently, the probability to satisfy the isolation constraint will increase and the feasibility region of the optimization problem is expanded. When the number of cell-edge and central users in the network are equal, i.e., point $50\%$ in Fig. \ref{fig:densityuti}, our proposed scheme achieves $15\%$ and $20\%$ higher utility performance compared to the OMA and NOMA cases, respectively, due to its capability to choose the best technology depending on users CSI and cost of technology implementation.

From Fig. \ref{fig:densityrate}, it is evident that all access methods have similar performance in terms of total rate of users when all users are in the central region. Comparing two points of $50\%$ and $80\%$ in Fig. \ref{fig:densityrate}, it can be observed that the gap between the total rate of our proposed scheme and NOMA decreases. In fact, when the number of users at cell edge and central region are equal, (i.e., point $50\%$), the flexible access technology selection of our proposed scheme leads to $17\%$ improvement in the total rate.



\begin{table}[]
\caption {Achieved utility of the proposed resource allocation algorithm and optimal resource allocation}
  \label{table}
\begin{tabular}{l|l|l|}
\cline{2-3}
& Exhaustive search for OMA/NOMA & Our proposed algorithm for OMA/NOMA  \\ \hline
\multicolumn{1}{|l|}{Achieved Utility ($N=3$, $K=6$)} & 30.5326 & 29.5603 \\ \hline
\multicolumn{1}{|l|}{Achieved Utility ($N=4$, $K=6$)} & 44.2093 & 41.0337 \\ \hline
\end{tabular}
\end{table}

To study the gap between our proposed scheme and the optimal resource allocation, we use the exhaustive search to find the optimal solution. First, we note that, the complexity of technology selection and user assignment of exhaustive search  is ${\Big( \bigg( \begin{matrix}K  \\2  \\\end{matrix} \bigg)+K \Big)^{N}}$. Consequently, the exhaustive search suffers from huge computational complexity at large values of $N$ and $K$. This forces us to focus on $K=6$ and $N=3 \textrm{ and } 4$ for this simulation. We also set $P_{\textrm{max}}=100$ W and $R_s=12$ bps/Hz. The results are summarized in Table \ref{table}, which reveals that the gap between the optimal resource allocation and our proposed algorithm is negligible when $N$ and $K$ are small, while this gap expands when network size grows. However, this gap does not exceed $7 \%$ as seen from Table \ref{table}.


\section{Conclusion}\label{sec:conclusion}
In this paper, we have introduced a new approach for access technology that can dynamically choose a suitable technology based on the instantaneous CSI. As an example, two different access technologies (OMA and NOMA) have been considered as options of access technology selection. We have considered a multi-user multi-carrier single cell downlink communication system, assuming a set of the SPs, each of which having a set of its own users and a minimum QoS requirement. We have then proposed a novel algorithm that can allocate resources including subcarriers, power and technology selection variables.
 We have defined a novel utility function which reflects the tradeoff between the achievable rate and the imposed cost for NOMA processing. To efficiently solve the proposed resource allocation problem, we have developed a two-step iterative algorithm. In the first step, by introducing auxiliary variables, the subcarrier assignment and technology selection problem is transformed and solved using linear integer programming. Subsequently, in the second step, the power allocation is solved by applying DC programming. Simulation results highlight that higher utility and lower outage probability can be achieved via the proposed dynamic access technology selection.

\appendices
\section{Appendix A: Power Allocation Strategy}\label{appen}

In the primal problem, we want to find the variables of matrix $\bm{P}$ with dimension $K\times N$ by considering fix value for $\bm{\lambda,\gamma,\zeta,\eta}$. Because each user can be first or second user of NOMA, we take derivative of general formula (\ref{equpowerlag}) with respect to $p_{k',n}$
\begin{eqnarray}\label{equpowerdif}
\lefteqn{\frac{dL(\bm{P},\bm{\lambda,\gamma,\zeta,\eta})}{dp_{k',n}}=}\nonumber\\&&\alpha_{k',k',n}^t\frac{h_{k',n}}{\sigma^2+p_{k',n}h_{k',n}}+\sum_{k_2\in\mathcal{K},k_2\neq k'}\Big(\frac{h_{k_2,n}}{p_{k',n}h_{k_2,n}+p_{k_2,n}h_{k_2,n}+\sigma^2}
-\frac{h_{k_2,n}}{p_{k',n}^{t_2-1}h_{k_2,n}+\sigma^2}\Big)u_{k',k_2,n}^t\nonumber\\&&
+\sum_{k_1\in\mathcal{K},k_1\neq k'}\frac{h_{k',n}u_{k_1,k',n}^t}{p_{k_1,n}h_{k',n}+p_{k',n}h_{k',n}+\sigma^2}\nonumber\\&&-V\Big(\sum_{k_1\in \mathcal{K},k_1\neq k'}u_{k_1,k',n}^t
\frac{h_{k',n}}{p_{k',n}^{t_2-1}h_{k',n}+\sigma^2}
-\sum_{k_2\in\mathcal{K},k_2\neq k'}u_{k',k_2,n}^t\frac{1}{p_{k',n}}\Big)
\nonumber\\&&+\lambda_{s'}(\alpha_{k',k',n}^t\frac{h_{k',n}}{\sigma^2+p_{k',n}h_{k',n}}+\underbrace{\sum_{k_1\in\mathcal{K},k_1\neq k'}\frac{h_{k',n}}{p_{k',n}h_{k',n}+p_{k_1,n}h_{k',n}+\sigma^2}u_{k_1,k',n}^t}_{\textrm{eighth term}})
\nonumber\\&&
+\sum_{s\in \mathcal{S},s\neq s'}\lambda_{s}\sum_{k_2\in\mathcal{K_{\textrm{s}}}}u_{k',k_2,n}^t\left(\frac{h_{k_2,n}}{p_{k_2,n}h_{k_1,n}+p_{k',n}h_{k_2,n}+\sigma^2}-\frac{h_{k_2,n}}{p_{k',n}^{t_2-1}h_{k_2,n}+\sigma^2}\right)
\nonumber
\end{eqnarray}
\begin{eqnarray}\label{equpowerdif}
+\gamma_n\left(\frac{-h_{k',n}\beta_n^t\alpha_{k',k',n}^t}{\sigma^2}+\!\!\!\!\!\sum_{k_1\in\mathcal{K},k_1\neq k'}\!\!\!\!\!\!u_{k_1,k',n}^t\frac{h_{k_1,n}}{\sigma^2}\right)-\zeta_{k',n}
-\eta\left(\alpha_{k',k',n}^t+\sum_{k_1\in\mathcal{K}}u_{k_1,k,n}^t\right),~\forall k',n.\nonumber\\
\end{eqnarray}

In (\ref{equpowerdif}), the first three terms are for the case where user $k'$ is the first user of NOMA or OMA user. Note that the second and third terms are in the case where the power of this user is interference for other users. In \eqref{equpowerdif}, since we do not replace $\alpha_{k',k',n}^t$ with its derived values from previous step (considering general format), the summation over $k_1$ appears when the user $k'$ is the second user of NOMA. 
	 Again, when user $k'$ is the second user of NOMA and it is belong to $s'$ SP, its rate should be considered in rate constraint of $s'$ SP which leads to the eighth term.

 All terms of $\frac{dL}{dp_{k',n}}$ that are not function of $p_{k,n},k=1,...,K$, in the above equation are represented by $D_{k',n}$ as,
  \begin{eqnarray}\label{equpowerd}
D_{k',n}\!\!\!&=\!\!\!&\beta_n^t\sum_{k_2,k_2\neq k'}\Big(
-\frac{h_{k_2,n}}{p_{k',n}^{t_2-1}h_{k_2,n}+\sigma^2}\Big)u_{k',k_2,n}^t
-V\Big(\sum_{k_1\in \mathcal{K},k_1\neq k'}u_{k_1,k',n}^t
\frac{h_{k',n}}{p_{k',n}^{t_2-1}h_{k',n}+\sigma^2}\Big)\\&&+
\sum_{s\in \mathcal{S},s\neq s'}\lambda_{s}\sum_{k_2\in\mathcal{K_{\textrm{s}}}}u_{k',k_2,n}^t\left(-\frac{h_{k_2,n}}{p_{k',n}^{t_2-1}h_{k_2,n}+\sigma^2}\right)
\nonumber\\&&
+\gamma_n\left(\frac{-h_{k',n}\beta_n^t\alpha_{k',k',n}^t}{\sigma^2}+\sum_{k_1,k_1\neq k'}u_{k_1,k',n}^t\frac{h_{k_1,n}}{\sigma^2}\right)-\zeta_{k',n}
-\eta\left(\alpha_{k',k',n}^t+\sum_{k_1}u_{k_1,k',n}^t\right)\nonumber
\end{eqnarray}

 Therefore, the relation between the optimum values of $p_{k,n}, k=1,..,K$ is
\begin{eqnarray}\label{equpowerkol}
&&\alpha_{k',k',n}^t\frac{h_{k',n}}{p_{k',n}h_{k',n}+\sigma^2}(1+\lambda_{s'})+\sum_{k,k\neq k'}\frac{h_{k,n}}{p_{k',n}h_{k,n}+p_{k,n}h_{k,n}+\sigma^2}u_{k',k,n}^t(1+\sum_{s\in S,k\in\mathcal{K_{\textrm{s}}}}\lambda_{s})\nonumber\\&&
+\sum_{k,k\neq k'}\frac{h_{k',n}u_{k,k',n}^t}{p_{k,n}h_{k',n}+p_{k',n}h_{k',n}+\sigma^2}(1+\lambda_{s'})+V\Big(\sum_{k,k\neq k'}u_{k',k,n}^t\frac{1}{p_{k',n}}\Big)+D_{k',n}=0\nonumber\\
\end{eqnarray}
Note that by allocation in step 1 and considering only two users for NOMA at most, there are two equations and two variables for each subcarrier. Next, we consider two cases.

In the first case, OMA is selected for subcarrier $n$ and the user $k$ uses it. Thus, $\beta_n^t=0$, $u_{k_1,k_2,n}^t=0,~\forall k_1,k_2$ and $\alpha_{k,k,n}^t=1$. So, by substituting these value in (\ref{equpowerkol}), the allocated power to the user $k$ on subcarrier $n$ is obtained according to (\ref{equpoweroma}).

 For subcarrier $n$, when NOMA is selected and if user $k_1$ and $k_2$ are  the first and second users, respectively, we have $\beta_n^t=1$, $\alpha_{k_1,k_1,n}^t=1$, $u_{k_1,k_2,n}^t=1$, and other variables, i.e., $\bm{\lambda,\gamma,\zeta,\eta}$, are equal to zero. By assuming $p_n=p_{k_1,n}+p_{k_2,n}$ we have two equations and two variables
 \begin{eqnarray}\label{equpower}
&&\frac{h_{k_1,n}}{p_{k_1,n}h_{k_1,n}+\sigma^2}(1+\lambda_{s'})+\frac{h_{k_2,n}}{p_{n}h_{k_2,n}+\sigma^2}(1+\lambda_{s''})+\frac{V}{p_{k_1,n}}+D_{k_1,n}=0\\
&&\frac{h_{k_2,n}}{p_{n}h_{k_2,n}+\sigma^2}(1+\lambda_{s'})+D_{k_2,n}=0
\end{eqnarray}
where user $k_1$ and $k_2$ are in $\mathcal{K}_{s'}$ and $\mathcal{K}_{{s''}}$, respectively. The quadratic equation for obtaining $p_{k_1,n}$ is as
 \begin{eqnarray}\label{equpower}
&&(\frac{-{{D}_{k_2,n}}(1+{{\lambda }_{s''}})}{1+{{\lambda }_{s'}}}+{{D}_{k_1,n}}){{h}_{k_1,n}}p_{k_1,n}^{2}+({{h}_{k_1,n}}(1+{{\lambda }_{s'}})+{\sigma^2}(\frac{-{{D}_{k_2,n}}(1+{{\lambda }_{s''}})}{1+{{\lambda }_{s'}}}+{{D}_{k_1,n}})\nonumber\\&&+V{{h}_{k_1,n}}){{p}_{k_1,n}}+V{\sigma^2}=0
\end{eqnarray}
As a result, the power of user $k_1$ and $k_2$ on the subcarrier $n$ are obtained by (\ref{equpowernoma}).

\bibliographystyle{IEEE}
\bibliography{thesis-bib3}

\begin{thebibliography}{10}

\bibitem{exponen1}
{Report ITU-R M.2370-0},
\newblock ``\textsc{IMT} traffic estimates for the years 2020 to 2030,''
\newblock {\em Technical report}, 2008.

\bibitem{exponen2}
{I. Cisco},
\newblock ``Cisco visual networking index: Forecast and methodology,
  2011–2016,''
\newblock {\em CISCO White paper}, 2012.

\bibitem{sad1}
{M. Mozaffari, W. Saad, M. Bennis, and M. Debbah},
\newblock ``Unmanned aerial vehicle with underlaid device-to-device
  communications: Performance and tradeoffs,''
\newblock {\em IEEE Transactions on Wireless Communications}, vol. 15, no. 6,
  pp. 3949--3963, Jun. 2016.

\bibitem{nomafirst}
{Y. Saito, Y. Kishiyama, A. Benjebbour, T. Nakamura, A. Li, and K. Higuchi},
\newblock ``Non-orthogonal multiple access (\textsc{NOMA}) for cellular future
  radio access,''
\newblock in {\em Proc. IEEE Vehicular Technology Conference}, Dresden,
  Germany, Jun. 2013, pp. 1--5.

\bibitem{sadA}
{M. El-Bamby, M. Bennis, W. Saad, M. Debbah, and M. Latva-aho},
\newblock ``Resource optimization and power allocation in full duplex
  non-orthogonal multiple access (\textsc{FD-NOMA}) networks,''
\newblock {\em IEEE Journal on Selected Areas in Communications}, vol. 35, no.
  12, pp. 2860--2873, Dec. 2017.

\bibitem{sadB}
{T. Park, G. Lee, and W. Saad},
\newblock ``Message-aware uplink transmit power level partitioning for
  non-orthogonal multiple access (\textsc{NOMA}),''
\newblock in {\em IEEE Global Communications Conference (GLOBECOM), Next
  Generation Networking and Internet Symposium}, Abu Dhabi, Dec. 2018.

\bibitem{scma}
{ H. Nikopour and H. Baligh},
\newblock ``Sparse code multiple access,''
\newblock in {\em in Proc. IEEE Int. Symp. Personal, Indoor, and Mobile Radio
  Commun. (PIMRC)}, London, UK, Sept. 2013, pp. 332–--336.

\bibitem{pdma}
{S. Chen, B. Ren, Q. Gao, S. Kang, S. Sun, and K. Niu},
\newblock ``Pattern division multiple access (\textsc{PDMA}) - \textsc{A} novel
  non-orthogonal multiple access for 5\textsc{G} radio networks,''
\newblock {\em IEEE Transactions on Vehicular Technology}, vol. 66, no. 4, pp.
  3185--3196, Jul. 2016.

\bibitem{secnoma}
{Y. Saito, A. Benjebbour, Y. Kishiyama, and T. Nakamura},
\newblock ``System level performance evaluation of downlink nonorthogonal
  multiple access (\textsc{NOMA}),''
\newblock in {\em Proc. IEEE Int. Symposium on Personal, Indoor and Mobile
  Radio Commun.}, London, UK, Sept. 2013.

\bibitem{advannoma2}
{W. Han, J. Ge, J. Men},
\newblock ``Performance analysis for \textsc{NOMA} energy harvesting relaying
  networks with transmit antenna selection and maximal-ratio combining over
  \textsc{N}akagami-m fading,''
\newblock {\em IET Communication}, vol. 10, no. 18, pp. 2687--2693, Dec. 2016.

\bibitem{advannoma3}
{B. He, A. Liu, N. Yang, V. K. N. Lau},
\newblock ``On the design of secure non-orthogonal multiple access systems,''
\newblock {\em IEEE Journal on Selected Areas in Communications}, vol. 35, no.
  10, pp. 2196--2206, Jul. 2017.

\bibitem{advannoma}
{S. Islam, M. Zeng, O. A. Dobre, and K.-S. Kwak},
\newblock ``Resource allocation for downlink \textsc{NOMA} systems: Key
  techniques and open issues,''
\newblock {\em IEEE Wireless Communications}, vol. 25, no. 2, pp. 40--47, Apr.
  2018.

\bibitem{csinoma}
{Z. Wei, D.W.K. Ng, J. Yuan, and H.M. Wang},
\newblock ``Optimal resource allocation for power-efficient \textsc{MC-NOMA}
  with imperfect channel state information,''
\newblock {\em IEEE Transactions on Communications}, vol. 65, no. 9, pp.
  3944--3961, Sept. 2017.

\bibitem{sdr}
{H. H. Cho},
\newblock ``Integration of \textsc{SDR} and \textsc{SDN} for 5\textsc{G},''
\newblock {\em IEEE Access}, vol. 2, pp. 1196--1204, Sept. 2014.

\bibitem{lte}
{H. Ekstr¨om, A. Furusk¨ar, J. Karlsson, M. Meyer, S. Parkvall, J. Torsner,
  and M. Wahlqvist},
\newblock ``Technical solutions for the 3\textsc{G} long-term evolution,''
\newblock {\em IEEE Communications Magazine}, vol. 44, no. 3, pp. 38--45, Mar.
  2006.

\bibitem{virtu}
{H. Wen, P. K. Tiwary, and T. Le-Ngoc},
\newblock {\em Wireless virtualization},
\newblock Springer, Cham, 2013.

\bibitem{update}
{D. Bertsekas},
\newblock {\em Nonlinear Programming},
\newblock 2nd ed. , Belmont, Massachusetts: Athena Scientific, 1999.

\bibitem{dc}
{S. Boyd, L. Xiao, A. Mutapic, and J. Mattingley},
\newblock ``Sequential convex programming,''
\newblock in {\em Stanford University, Stanford, CA, USA Technical Report
  EE364b}, 2007.

\bibitem{omasur}
{S. Sadr, A. Anpalagan, and K. Raahemifar},
\newblock ``Radio resource allocation algorithms for the downlink of multiuser
  \textsc{OFDM} communication systems,''
\newblock {\em IEEE Communications Surveys and Tutorials}, vol. 11, no. 3, pp.
  92–--106, Aug. 2009.

\bibitem{omasur2}
{A.S. Hamza, S.S. Khalifa, H.S. Hamza},
\newblock ``A survey on inter-cell interference coordination techniques in
  \textsc{OFDMA}-based cellular networks,''
\newblock {\em IEEE Communications Surveys and Tutorials}, vol. 15, no. 4, pp.
  1642--1670, Fourth \textsc{Q}uarter 2013.

\bibitem{omasur3}
{M. Naeem, A. Anpalagan, M. Jaseemuddin, and D. C. Lee },
\newblock ``Resource allocation techniques in cooperative cognitive radio
  networks,''
\newblock {\em IEEE Communications Surveys and Tutorials}, vol. 16, no. 2, pp.
  729--744, Nov. 2014.

\bibitem{omasur4}
{R. Afolabi, A. Dadlani, and K. Kim},
\newblock ``Multicast scheduling and resource allocation algorithms for
  \textsc{OFDMA}-based systems: A survey,''
\newblock {\em IEEE Communications Surveys and Tutorials}, vol. 15, no. 1, pp.
  240–--254, Feb. 2013.

\bibitem{nphard}
{L. Lei, D. Yuan, C. K. Ho, and S. Sun},
\newblock ``Power and channel allocation for non-orthogonal multiple access in
  5\textsc{G} systems: Tractability and computation,''
\newblock {\em IEEE Transactions on Wireless Communications}, vol. 15, no. 12,
  pp. 8580--–8594, Dec. 2016.

\bibitem{ratenoma}
{B. Di, L. Song, and Y. Li},
\newblock ``Sub-channel assignment, power allocation, and user scheduling for
  non-orthogonal multiple access networks,''
\newblock {\em IEEE Transactions on Wireless Communications}, vol. 15, no. 11,
  pp. 7686–--7698, Nov. 2016.

\bibitem{eenoma}
{F. Fang, H. Zhang, J. Cheng, and V. C. M. Leung},
\newblock ``Energy-efficient resource allocation for downlink non-orthogonal
  multiple access network,''
\newblock {\em IEEE Transactions on Communications}, vol. 64, no. 9, pp.
  3722–--3732, Sept. 2016.

\bibitem{varobj}
{J. Zhu, J. Wang, Y. Huang, S. He, X. You, and L. Yang},
\newblock ``On optimal power allocation for downlink non-orthogonal multiple
  access systems,''
\newblock {\em IEEE Journal on Selected Areas in Communications}, vol. 35, no.
  12, pp. 2744--2757, Jul. 2017.

\bibitem{lowcom}
{J. He, and Z. Tang},
\newblock ``Low-complexity user pairing and power allocation algorithm for
  5\textsc{G} cellular network non-orthogonal multiple access,''
\newblock {\em Electronics Letters}, vol. 53, no. 9, pp. 626--627, May 2017.

\bibitem{npofdm}
{C. Y. Wong, R. S. Cheng, K. B. Letaief, and R. D. Murch},
\newblock ``Multiuser \textsc{OFDM} with adaptive subcarrier, bit, and power
  allocation,''
\newblock {\em IEEE Journal on Selected Areas in Communications}, vol. 17, no.
  10, pp. 1747--1758, Oct. 1999.

\bibitem{duplexnoma}
{Y. Sun, D. W. K. Ng, Z. Ding, and R. Schober},
\newblock ``Optimal joint power and subcarrier allocation for full-duplex
  multicarrier non-orthogonal multiple access systems,''
\newblock {\em IEEE Transactions on Communications}, vol. 65, no. 3, pp.
  1077--–1091, Mar. 2017.

\bibitem{mimonoma}
{S. Ali, E. Hossain, and D.I. Kim},
\newblock ``Non-orthogonal multiple access (\textsc{NOMA}) for downlink
  multiuser \textsc{MIMO} systems: User clustering, beamforming, and power
  allocation,''
\newblock {\em IEEE Access}, vol. 5, pp. 565--577, Dec. 2017.

\bibitem{het}
{J. Zhao, Y. Liu, K. K. Chai, A. Nallanathan, Y. Chen, and Z. Han},
\newblock ``Low-complexity user pairing and power allocation algorithm for
  5\textsc{G} cellular network non-orthogonal multiple access,''
\newblock {\em IEEE Transactions on Wireless Communications}, vol. 16, no. 9,
  pp. 5825--5837, May 2017.

\bibitem{new}
{ A. J. Morgado, K. M. S. Huq, J. Rodriguez, C. Politis, and H. Gacanin},
\newblock ``Hybrid resource allocation for millimeter-wave \textsc{NOMA},''
\newblock {\em IEEE Wireless Communications Magazine}, vol. 24, no. 5, pp.
  23--29, Oct. 2017.

\bibitem{new2}
{A. S. Marcano, and H. L. Christiansen},
\newblock ``Impact of \textsc{NOMA} on network capacity dimensioning for
  5\textsc{G} \textsc{H}et\textsc{N}ets,''
\newblock {\em IEEE Access}, vol. 6, pp. 13587--13603, Feb. 2018.

\bibitem{access}
{M. Baghani, S. Parsaeefard, and T. Le-Ngoc},
\newblock ``Multi-objective resource allocation in density-aware design of
  \textsc{C-RAN} in 5\textsc{G},''
\newblock {\em IEEE Access}, vol. 6, pp. 45177--45190, Aug. 2018.

\bibitem{compnoma}
{L. Dai, B. Wang, Y. Yuan, S. Han, C.-L. I, Z. Wang},
\newblock ``Non-orthogonal multiple access for 5\textsc{G}: Solutions
  challenges opportunities and future research trends,''
\newblock {\em IEEE Communications Magazine}, vol. 53, no. 9, pp. 74--81, Sept.
  2015.

\bibitem{twousern}
{Z. Ding, P. Fan, H. V. Poor},
\newblock ``Impact of user pairing on 5\textsc{G} nonorthogonal multiple-access
  downlink transmissions,''
\newblock {\em IEEE Transactions on Vehicular Technology}, vol. 65, no. 8, pp.
  6010--6023, Aug. 2016.

\bibitem{complexreci}
{B. Ling, C. Dong, J. Dai, and J. Lin},
\newblock ``Multiple decision aided successive interference cancellation
  receiver for \textsc{NOMA} systems,''
\newblock {\em IEEE Wireless Communications Letter}, vol. 6, no. 4, pp.
  498--501, May 2017.

\bibitem{harq}
{ A. Li, A. Benjebbour, X. Chen, H. Jiang, and H. Kayama},
\newblock ``Investigation on hybrid automatic repeat request (\textsc{HARQ})
  design for \textsc{NOMA} with \textsc{SU-MIMO},''
\newblock in {\em IEEE 26th Annu. Int. Symp. Pers., Indoor, Mobile Radio
  Commun. (PIMRC)}, Hong Kong, China, Sept. 2015, pp. 590–--594.

\bibitem{MM}
{A. Taleb Zadeh Kasgari and W. Saad},
\newblock ``Stochastic optimization and control framework for 5\textsc{G}
  network slicing with effective isolation,''
\newblock in {\em IEEE 52nd Annual Conference on Information Sciences and
  Systems (CISS)}, Princeton, NJ, USA, Mar. 2018.

\bibitem{ratecons}
{R. Kokku, R. Mahindra, H. Zhang, and S. Rangarajan},
\newblock ``\textsc{NVS}: A substrate for virtualizing wireless resources in
  cellular networks,''
\newblock {\em Society for Industrial and Applied Mathematics Journal on
  Optimization}, vol. 20, no. 5, pp. 1333–--1346, Oct. 2012.

\bibitem{tolcons}
{M. S. Ali, H. Tabassum, and E. Hossain},
\newblock ``Dynamic user clustering and power allocation in non-orthogonal
  multiple access (\textsc{NOMA}) systems,''
\newblock {\em IEEE Access}, vol. 4, pp. 6325--6343, Aug. 2016.

\bibitem{cvxlin}
{M. Grant and S. Boyd},
\newblock ``\textsc{CVX}: \textsc{MATLAB} software for disciplined convex
  programming,''
\newblock in {\em Version 2.1. [Online]. Available: http://cvxr.com/cvx}, Mar.
  2017.

\bibitem{razavi}
{M. Razaviyayn, M. Hong, and Z. Q. Lue},
\newblock ``A unified convergence analysis of block successive minimization
  methods for nonsmooth optimization,''
\newblock {\em Society for Industrial and Applied Mathematics Journal on
  Optimization}, vol. 23, no. 2, pp. 1126–--1153, Jun. 2013.

\bibitem{LP}
{S. J. Wright},
\newblock {\em Primal-Dual Interior-Point Methods},
\newblock PA, Philadelphia:SIAM, 1997.

\bibitem{complex}
{B. Borchers and J. Young},
\newblock ``Implementation of a primal-dual method for \textsc{sdp} on a shared
  memory parallel architecture,''
\newblock {\em Computational Optimization and Applications}, vol. 37, no. 3,
  pp. 355--–369, Jul. 2007.

\bibitem{dccompl}
{ A. L. Yuille and A. Rangarajan},
\newblock ``The concave-convex procedure,''
\newblock {\em Neural Computation}, vol. 15, no. 4, pp. 915–--936, Apr. 2003.

\end{thebibliography}

\end{document}